\documentclass[prb,aps,showpacs,prb,amsmath,floatfix,twocolumn]{revtex4}
\usepackage[]{graphicx}
\usepackage{graphicx}
\usepackage{dcolumn}
\usepackage{bm}

\begin{document}
\title{Time dependent quantum transport through
Kondo correlated quantum dots}
\author{A. Goker$^{1}$ and E. Gedik$^{2}$}

\affiliation{$^1$
Department of Physics, \\
Bilecik University, \\
11210, G$\ddot{u}$l$\ddot{u}$mbe, Bilecik, Turkey
}

\affiliation{$^2$
Department of Physics, \\
Eskisehir Osmangazi University, \\
26480, Meselik, Eskisehir, Turkey
}

\date{\today}

\begin{abstract}
In this chapter, we review recent work about time
dependent quantum transport through a quantum
dot in Kondo regime. This represents a major step
towards designing next generation transistors that 
are expected to replace current MOSFET's in a few years.
We first discuss the effects of the density of states 
of gold contacts on the instantaneous conductance 
of an asymmetrically coupled quantum dot that is 
abruptly moved into Kondo regime via a gate voltage.
Next, we investigate the effect of strong electron-phonon
coupling on the dot on the instantaneous conductance. 
Finally, we discuss thermoelectric effects using linear 
response Onsager relations for a quantum dot that is 
either abruptly moved into Kondo regime or driven 
sinusoidally via a gate voltage. We explain
encountered peculiarities in transport 
based on the behaviour of the density of states 
of the dot and the evolution of the Kondo resonance.
 
\end{abstract}



\thispagestyle{headings}

\maketitle

\section{Introduction}

Single electron transistors are nanodevices made up of
a discrete energy level sandwiched between electrodes.
They are expected to have profound implications in 
designing next generation transistors since 
conventional MOSFET's will reach their
physical limit at 10 nm in a few years \cite{Likharev03}.
The rapid evolution of this field derived 
extra benefit from unprecedented control 
over nanodevices brought forward by nanotechnology
revolution. It is imperative that the
switching behaviour of these next generation
transistors are examined carefully before
they are integrated into electrical circuits. 
Furthermore, the field of time dependent
quantum transport through nanostructures 
should also help calibrate quantum computers
\cite{ElzermanetAl04Nature} and
single electron guns \cite{FeveetAl07Science}.

Several time scales emerged \cite{PlihaletAl05PRB,IzmaylovetAl06JPCM}
after the initial studies about the effect
of a strongly correlated state called Kondo 
effect on the transient current ensuing 
after a sudden shift of the gate or bias 
voltage \cite{NordlanderetAl99PRL,PlihaletAl00PRB,MerinoMarston04PRB}. 
In section~\ref{sec:designer} of this chapter, we will 
investigate the damped oscillations in the 
long timescale for an asymmetrically coupled system 
\cite{GokeretAl07JPCM,GokeretAl10PRB,GokeretAl11CPL}
arising as a result of strong interference
between the Kondo resonance and the Van Hove 
singularities in the contacts' density of states.

In section ~\ref{sec:vibrating}, we will consider 
the effect of strong electron-phonon coupling
on the instantaneous current through a single 
electron transistor abruptly moved into Kondo 
regime via a gate voltage. We will find out that 
the instantaneous current exhibits damped sinusoidal 
oscillations in the long timescale even in infinitesimal 
bias \cite{Goker11JPCM}. We will elaborate on
the origin of this phenomena based on the 
behaviour of the density of states of the dot.

Thermopower (Seebeck coefficient) can provide clue 
about the alignment of the dot level with respect
to the Fermi level of the contacts hence it 
is a valuable tool to determine the existence
of Kondo resonance. This has been verified 
with several recent experiments as well 
\cite{Reddyetal07Science,Bahetietal08NL,Malenetal09NL,TangetAl10APL}.

Different groups pointed out that the sign 
of thermopower in Kondo regime depends 
quite sensitively on position of the dot level
\cite{DongetAl02JPCM,CostietAl10PRB}. In 
section ~\ref{sec:thermal}, we will summarize 
recent efforts to study the transient behaviour of 
thermopower. We will start by discussing the behaviour 
of thermopower when the the dot energy level is 
suddenly shifted to a position near the Fermi 
level \cite{Goker2012}. We will show that the 
inverse of the saturated decay time of thermopower 
to its steady state value is equal to the Kondo 
temperature.

Time dependent ac perturbations of the dot 
energy level allows monitoring instantaneous
thermopower when the Kondo temperature changes 
constantly due to the sinusoidal motion of the
dot level \cite{AlhassaniehetAl05PRL,KiselevetAl06PRB}.
Experimental developments on adjusting
dot-lead coupling enabled this so called
Kondo shuttling \cite{ScheibleetAl04PRL,ParksetAl07PRL}.
Previously, it was found that the time averaged 
conductance deviates significantly from its monotonic 
decrease as a function of applied bias when the bias 
is equal to the driving frequency of the dot 
level \cite{Goker08SSC}. In the latter part of
section ~\ref{sec:thermal}, we will investigate 
in detail the instantaneous and time averaged 
values of the thermopower when the dot level is 
driven sinusoidally by means of a gate voltage
\cite{Gokeretal2012}.

Before we move on to discuss the results, we would like to
outline the theoretical framework for our calculations.
We will model the single electron transistor with 
a single discrete degenerate energy level $\varepsilon_{dot}$ 
coupled to the electrodes. Anderson Hamiltonian sufficiently
describes the physics of this set-up. It is given by
\begin{eqnarray}
H(t)&=&\sum_{\sigma} \varepsilon_{dot}(t)n_{\sigma}
+\sum_{k\alpha\sigma}\varepsilon_{k}n_{k\alpha\sigma}
+{\textstyle\frac{1}{2}}{\sum} U_{\sigma,\sigma'}n_{\sigma}n_{\sigma'}+ \nonumber \\
& &\sum_{k\alpha\sigma} \left[ V_{\alpha}(\varepsilon_{k\alpha},t)c_{k\alpha\sigma}^{\dag}c_{\sigma}+
{\rm H.c.} \right],
\label{Anderson}
\end{eqnarray}
where $c^\dagger_\sigma$ ($c_\sigma$) and 
$c^\dagger_{k\alpha\sigma}$ ($c_{k\alpha\sigma}$) with 
$\alpha$=L,R create (annihilate) an electron of spin 
$\sigma$ in the dot energy level and in the left(L) and 
right(R) electrodes respectively. The $n_\sigma$ and 
$n_{k\alpha\sigma}$ are the number operators for the dot 
level and the electrode $\alpha$. $V_{\alpha}$ are the 
tunneling amplitudes between the electrode $\alpha$ and 
the quantum dot. The Coulomb repulsion $U$ is taken to be 
infinite. This choice restricts the occupation of 
the dot level to unity. In this chapter, we will adopt 
atomic units where $\hbar=k_{\rm B}=e=1$.

In our method, slave boson transformation is carried out 
for the Anderson Hamiltonian. In this procedure, ordinary 
electron operators on the dot are expressed in terms of a 
massless (slave) boson operator and a pseudofermion 
operator as
\begin{equation}
c_{\sigma}=b^{\dagger} f_{\sigma},
\end{equation}
subject to 
\begin{equation}
Q=b^{\dagger}b+\sum_{\sigma}f^{\dagger}_{\sigma}f_{\sigma}=1.
\end{equation}
This last requirement is needed to prevent
the double occupancy in the original Hamiltonian
when $U \rightarrow \infty$. This transformation 
enables drop the quartic Hubbard term while retaining 
its effects in other terms. The transformed Hamiltonian 
turns out to be
\begin{eqnarray}
H(t)&=&\sum_{\sigma}\epsilon_{dot}(t)n_{\sigma}+ \nonumber \\
& &\sum_{k\alpha\sigma}\left [\epsilon_{k}n_{k\alpha\sigma}+
V_{\alpha}(\varepsilon_{k\alpha},t)c_{k\alpha\sigma}^{\dag}
b^{\dag}f_{\sigma}+{\rm H.c.} \right],
\end{eqnarray}
where $f_{\sigma}^{\dag}(f_{\sigma})$ and $b^{dag}(b)$
with $\alpha$=L,R create(annihilate) an electron of spin 
$\sigma$ and a slave boson on the dot respectively.

Hopping matrix elements will be assumed to have no explicit
time dependency. Consequently, the coupling of the quantum
dot to the electrodes can be expressed as
$\Gamma_{L(R)}(\epsilon)=\bar{\Gamma}_{L(R)} \rho_{L(R)}(\epsilon)$,
where $\bar{\Gamma}_{L(R)}$ is a constant given by
$\bar{\Gamma}_{L(R)}=2\pi|V_{L(R)}(\epsilon_f)|^2$ and $ \rho_{L(R)}(\epsilon)$
is the density of states function.

The retarded Green function can then be redefined      
in terms of the slave boson and pseudofermion        
Green functions \cite{GokeretAl07JPCM} as           
\begin{eqnarray}
G^R(t,t_1) &=& -i\theta(t-t_1)[G^R_{pseudo}(t,t_1)B^<(t_1,t) \nonumber \\
& & +G^<_{pseudo}(t,t_1)B^{R}(t_1,t)].
\end{eqnarray}

It is a quite a cumbersome and highly nontrivial 
task to compute the values of these double time 
Green functions in real time. This is managed by solving
coupled integro-differential Dyson equations in a 
two-dimensional cartesian grid. The self-energies 
of the pseudofermion and slave boson are key
ingredients into these equations. We utilize
quite versatile non-crossing approximation(NCA) for
this purpose \cite{ShaoetAl194PRB,IzmaylovetAl06JPCM}. 
NCA discards higher order corrections in which 
propagators cross each other. Therefore, it has 
certain drawbacks such as giving unphysical results 
in finite magnetic fields and when the ambient 
temperature is an order of magnitude smaller than 
Kondo energy scale. These regimes will be avoided 
in this chapter. Main advantage of NCA is that it 
is a quite powerful technique to obtain accurate 
results for dynamical quantities. Once the Dyson 
equations are solved, their values are stored in a 
square matrix which is propagated in discrete steps 
in time in diagonal direction.

Kondo effect is considered a prime example
of many body physics occurring at low temperatures
as a result of the coupling between the net spin 
localized within the dot and the continuum electrons 
of a metal in its vicinity. Its main revelation is 
a sharp resonance formed around the Fermi level of 
the metal. The linewidth of this Kondo resonance is
on the order of a low energy scale called
the Kondo temperature which is denoted by $T_K$
and expressed as
\begin{equation}
T_K \approx \left(\frac{D\Gamma_{tot}}{4}\right)^\frac{1}{2}
\exp\left(-\frac{\pi|\epsilon_{\rm dot}|}{\Gamma_{tot}}\right),
\label{tkondo}
\end{equation}
In Eq.~(\ref{tkondo}) $D$ is the half bandwidth of the
conduction electrons of the electrode
whereas $\Gamma_{tot}=(\bar{\Gamma}_L+\bar{\Gamma}_R) \rho(\epsilon_f)$.

\section{Designer switches}
\label{sec:designer}

\begin{figure}[htbp]
\centerline{\includegraphics[angle=0,width=8.6cm,height=6.2cm]{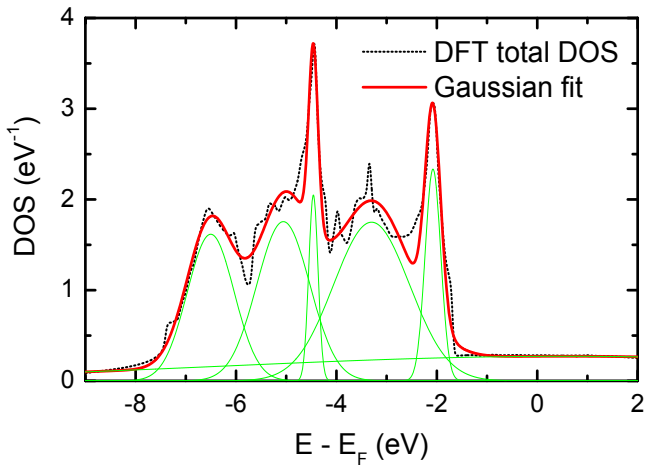}}
\caption{
Density of states of gold calculated using DFT is shown with black dashed curve
as a function of separation from the Fermi level. Red curve corresponds to the
best fit to the ab inito data by using a linear combination of Gaussian functions.
Each individual Gaussian function is also shown with green curves.
}
\label{PRB1}
\end{figure}

In this section, we will study the instantaneous 
conductance for a situation where the dot level is 
abruptly shifted from $\epsilon_1=-5\Gamma_{tot}$ to 
$\epsilon_2=-2\Gamma_{tot}$ where $\Gamma_{tot}=$0.8 eV
at $t=0$ by a gate voltage. Both contacts were assumed
to be made up of gold whose density of states is 
shown in Fig.~\ref{PRB1}. The density of states of
the gold contacts were obtained with the ab inito 
calculations which invoked the full-potential linearized augmented
plane wave (FP-LAPW) method using WIEN2K package
\cite{Blahaetal01Book}. The generalized
gradient approximation devised by Perdew, Burke, and
Ernzerhof (GGA-PBE) form \cite{Perdewetal96PRL} was
utilized as exchange-correlation functional 
in all ab initio calculations. The core
states of gold atom were taken to be in an electronic 
configuration of (Kr, $4d^{10}4f^{14}5s^{2}$).
The valence states include $5p$, $5d$, $6s$, and $6p$. Our
calculations invoked the unit cell of $fcc$ Au (space group
$Fm\bar{3}m$, $a$ = 4.080 \AA) with 4 atoms per cell.

We carried out a fitting procedure using six different
Gaussians with various linewidth and peak positions to 
mimic the density of states of gold. This enabled
us to express the density of states of gold in a 
functional form. The resulting best fit curves 
are also shown in red in Fig.~\ref{PRB1}.

The net current is obtained from the pseudofermion 
and slave boson Green's functions $G_{pseu}^{<(R)}(t,t')$ and
$B^{<(R)}(t,t')$. The general expression for the net 
current \cite{JauhoetAl94PRB} can be rewritten by using
these Green's functions \cite{GokeretAl07JPCM}.
The final expression is given by
\begin{eqnarray}
& & I(t) =-2(\bar{\Gamma}_{L}-\bar{\Gamma}_{R})\textit{Re} \left (\int_{-\infty}^{t} dt_1
\xi_{o}(t,t_1)h(t-t_1)\right)+\nonumber \\
& & 2\bar{\Gamma}_{L} Re \left (\int_{-\infty}^{t} dt_1
(\xi_{o}(t,t_1)+\xi_{u}(t,t_1)) f_{L}(t-t_1) \right)- \nonumber \\
& & 2\bar{\Gamma}_{R} Re \left (\int_{-\infty}^{t} dt_1
(\xi_{o}(t,t_1)+\xi_{u}(t,t_1)) f_{R}(t-t_1)\right)\nonumber \\
\label{current}
\end{eqnarray}
with $\xi_{o}(t,t_1)=G_{pseu}^{<}(t,t_1)B^{R}(t_1,t)$ and
$\xi_{u}(t,t_1)=G_{pseu}^{R}(t,t_1)B^{<}(t_1,t)$. In Eq.~(\ref{current}),
$f_L(t-t_1)$ and $f_R(t-t_1)$ are the convolution of the
density of states function with the Fermi-Dirac distributions of
the left and right contacts respectively, whereas
$h(t-t_1)$ is the Fourier transform of the DOS \cite{GokeretAl07JPCM}. The
conductance $G$ is equal to the current divided by the bias voltage
$V$. $\eta=\frac{\bar{\Gamma}_{L}}{\bar{\Gamma}_{tot}}$, where
$\bar{\Gamma}_{tot}=\bar{\Gamma}_{L}+\bar{\Gamma}_{R}$,
will be called the asymmetry factor.

\begin{figure}[htbp]
\centerline{\includegraphics[angle=0,width=8.4cm,height=6.0cm]{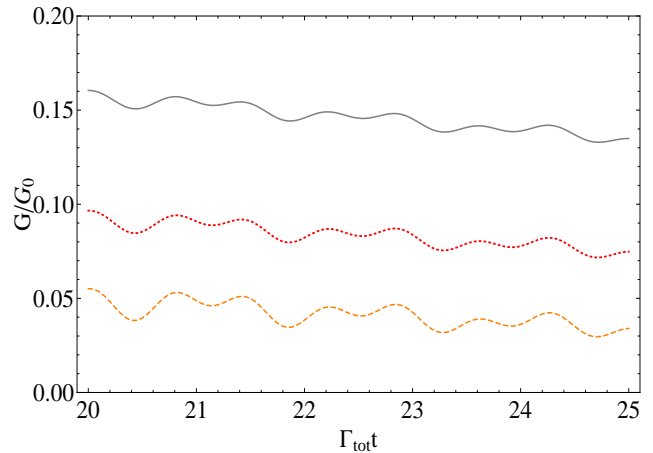}}
\caption{
Grey, red and orange curves from top to bottom show the instantaneous
conductance versus time in Kondo timescale after the dot level has been
switched to its final position for asymmetry factors of 0.95, 0.9 and 0.85
respectively at T=0.009$\Gamma_{tot}$ and V=$T_K$ for constant $\Gamma_{tot}$.
}
\label{PRB2}
\end{figure}

We assume that the dot level is switched to its
final position at $t$=0. In the initial short 
timescale corresponding to $\Gamma t \le$5, the conductance 
attains a maximum value for large asymmetry factors 
before starting to decay in agreement with previous 
studies \cite{GokeretAl07JPCM}. The instantaneous 
conductance in the long Kondo timescale is shown 
in Fig.~\ref{PRB2}. It is clear that a complex ringing 
is taking place here and the amplitude of the oscillations 
dwindles as the asymmetry factor is reduced. Oscillations
completely die out for symmetric coupling since the 
interference between the left contact and the Kondo 
resonance is out of phase with the interference between 
the right contact and the Kondo resonance due to opposite 
signs in Eq.~(\ref{current}). The amplitudes of two 
simultaneous interference effects are equal for 
symmetric coupling hence the oscillations cancel 
each other.

Fourier transform of the time dependent conductance
reveals all the frequency components causing
the complex fluctuation pattern. This analysis
provides two distinct frequencies, $\omega_1$ and 
$\omega_2$, which give rise to beating with envelope 
and carrier frequencies of $\omega_1-\omega_2$ and 
$\omega_1+\omega_2$, respectively. It turns out that 
$\omega_1=2.15\omega_2$ corresponding to the
ratio of the separation of the peaks at $-2.08$ eV and 
$-4.46$ eV in Fig.~\ref{PRB1} from the Fermi level. 
This outcome is a clear evidence that that the 
transient current in the Kondo regime is a tool
to detect the discontinuities in the density of 
states of the electrodes.

\begin{figure}[htbp]
\centerline{\includegraphics[angle=0,width=8.4cm,height=6.0cm]{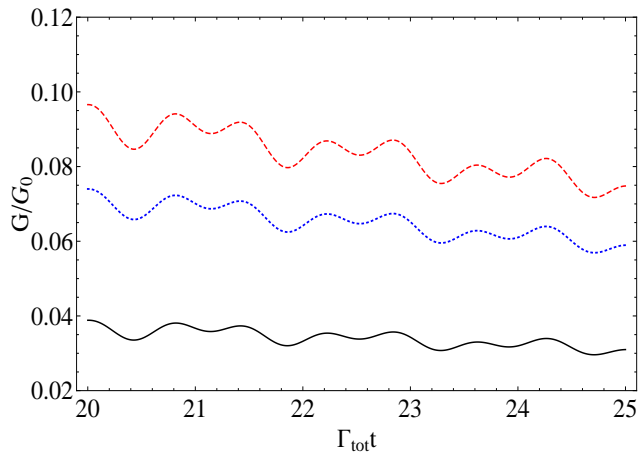}}
\caption{
Red, blue and black curves from top to bottom represent the
instantaneous conductance versus time in Kondo timescale after
the dot level has been switched to its final position for
asymmetry factor of 0.9 at T=0.009$\Gamma_{tot}$ when the
bias is equal to V=$T_K$, V=4$T_K$ and V=8$T_K$ respectively
for constant $\Gamma_{tot}$.
}
\label{PRB3}
\end{figure}

The influence of finite bias is displayed
in Fig.~\ref{PRB3}. Finite bias quenches the 
amplitude of oscillations. Moreover, the decay 
rate of these oscillations, defined to be the 
inverse of the time required for them to disappear,
increases. This is a clue about the relation of these
oscillations to the formation of the Kondo resonance.

\begin{figure}[htbp]
\centerline{\includegraphics[angle=0,width=8.4cm,height=6.0cm]{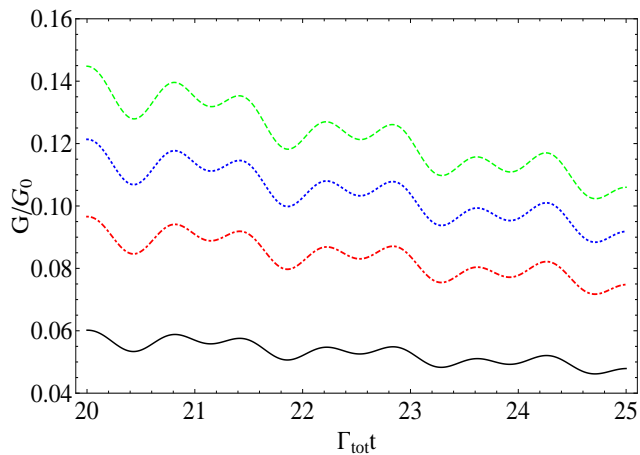}}
\caption{
Green, black, blue, red and purple curves from top to bottom show the
instantaneous conductance versus time in Kondo timescale after the dot
level has been switched to its final position for asymmetry factor of 0.9
at T=0.0015$\Gamma_{tot}$, T=0.0060$\Gamma_{tot}$, T=0.0090$\Gamma_{tot}$ 
and T=0.0150$\Gamma_{tot}$ respectively for constant $\Gamma_{tot}$ with a 
bias of V=$T_K$. 
}
\label{PRB4}
\end{figure}

Fig.~\ref{PRB4} demonstrates the effect of ambient 
temperature. The decay rate of the oscillations 
decreases upon reducing the ambient temperature. 
Furthermore, lower ambient temperatures increases 
the amplitude of the oscillations. However, it 
saturates when the temperature approaches the Kondo 
temperature $T_K$. Lowering the temperature below 
$T_K$ doesn't change the amplitude.

\begin{figure}[htbp]
\centerline{\includegraphics[angle=0,width=6.6cm,height=7.0cm]{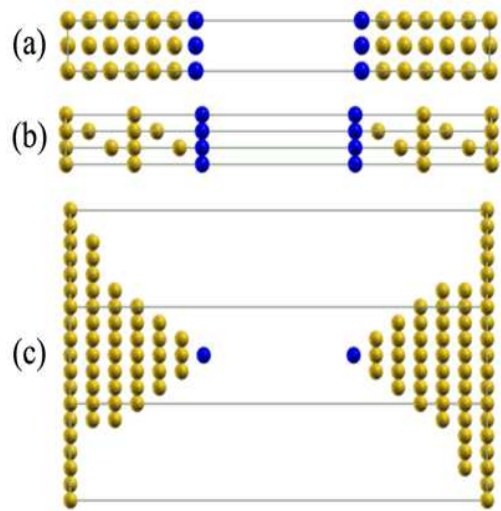}}
\caption{
This figure shows the three geometries, (a) (001)-surface,
(b) (111)-surface and (c) (111)-pyramide, used to simulate
the profile of Au electrode. The Au atoms in blue color are
the atoms relevant to the transport properties between
electrodes in all three structures.
}
\label{CPL1}
\end{figure}

\begin{figure}[htbp]
\begin{center}$
\begin{array}{c}
\includegraphics[angle=0,width=6.8cm,height=5.0cm]{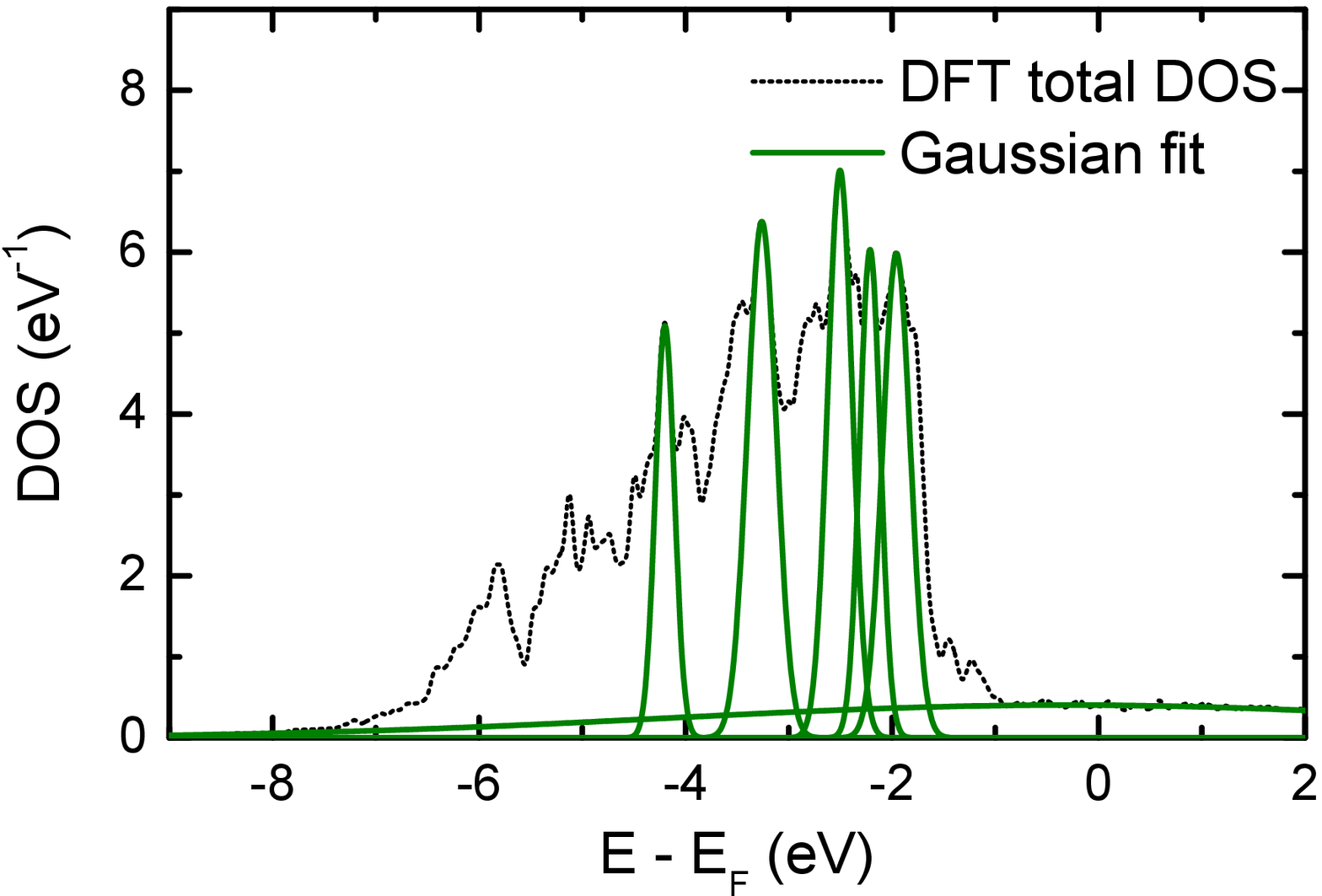} \\
\includegraphics[angle=0,width=6.8cm,height=5.0cm]{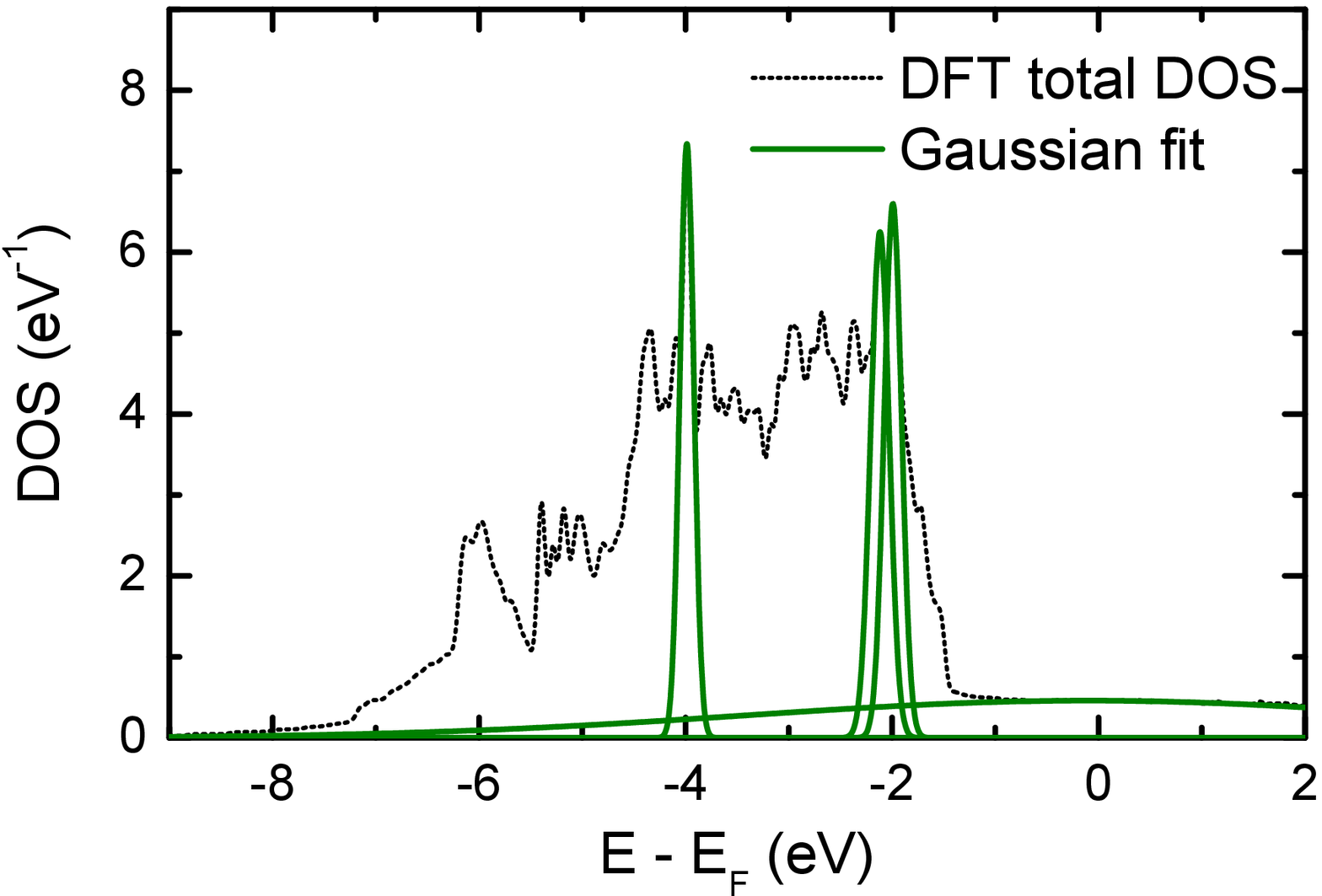} \\
\includegraphics[angle=0,width=6.8cm,height=5.0cm]{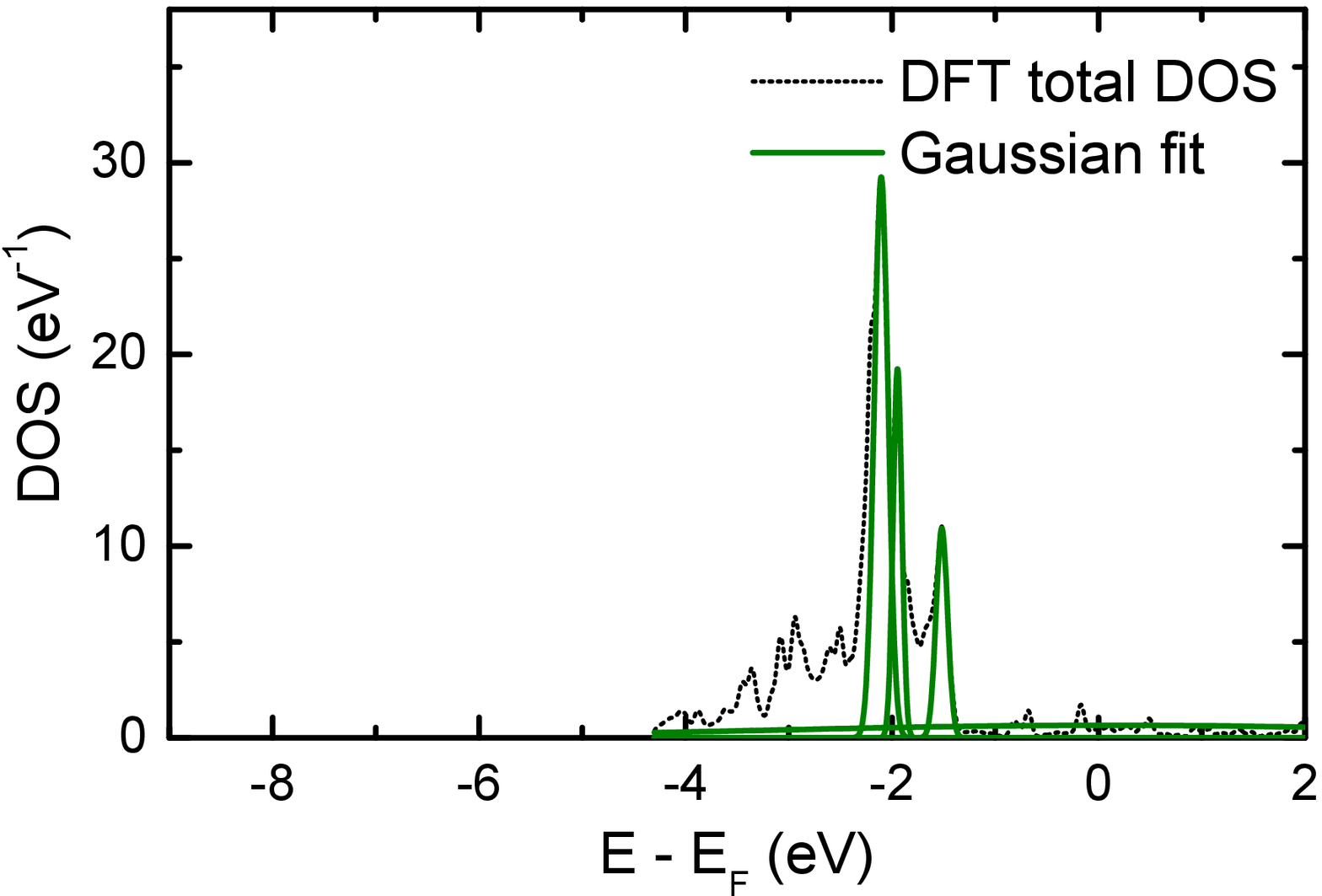}
\end{array}$
\end{center}
\caption{
Density of states of (001) surface, (111) surface and (111)
pyramid calculated using DFT is shown with black dashed curve
from top to bottom respectively as a function of separation
from the Fermi level. Each Gaussian used to capture the sharp
features and the Fermi level is also shown with green curves
in each geometry.
}
\label{CPL2}
\end{figure}

We now would like to analyze the effect of 
specific contact geometry on transient current.
To this end, we studied three geometries, 
i.e. (001)-surface, (111)-surface and (111)-pyramid
to simulate the Au electrode profile as shown 
in Fig.~\ref{CPL1}. Atomic slabs with 13 Au 
layers have been used to build these three 
geometries. The distance between the two opposite 
electrodes is 30 Bohr radius in all cases. The 
exchange-correlation potential of the generalized 
gradient approximation within the Perdew, Burke, and 
Ernzerhof (GGA-PBE) form \cite{Perdewetal96PRL} has been 
used in all calculations. The plane wave cut-off has been 
determined by $R_{mt}K_{max}$ = 6.5 and $l_{max}$ = 10. 
K-mesh of 30x30x3, 36x36x2 and 6x6x2 have been adopted 
for (001)-surface, (111)-surface and (111)-pyramid 
respectively. Note that only the DOS of topmost atoms, 
which are relevant to the transport properties between 
electrodes in the surface/pyramid structures, will be 
used in the following many body calculations. The 
resulting density of states is shown in Fig.~\ref{CPL2}.

We will be concerned with the transient current ensuing
after the dot level is switched from $\epsilon_1=-4\Gamma_{tot}$
to $\epsilon_2=-2\Gamma_{tot}$ at t=0 via a gate voltage.
For all three geometries, this triggers an abrupt shift from 
a non-Kondo state to a Kondo state. It is worth mentioning 
that the Kondo temperature in the final state is slightly 
lower for (111) pyramide than the other two geometries due 
to the shorter conduction electron bandwidth in Fig.~\ref{PRB1}.

\begin{figure}[htbp]
\begin{center}$
\begin{array}{c}
\includegraphics[angle=0,width=6.8cm,height=5.0cm]{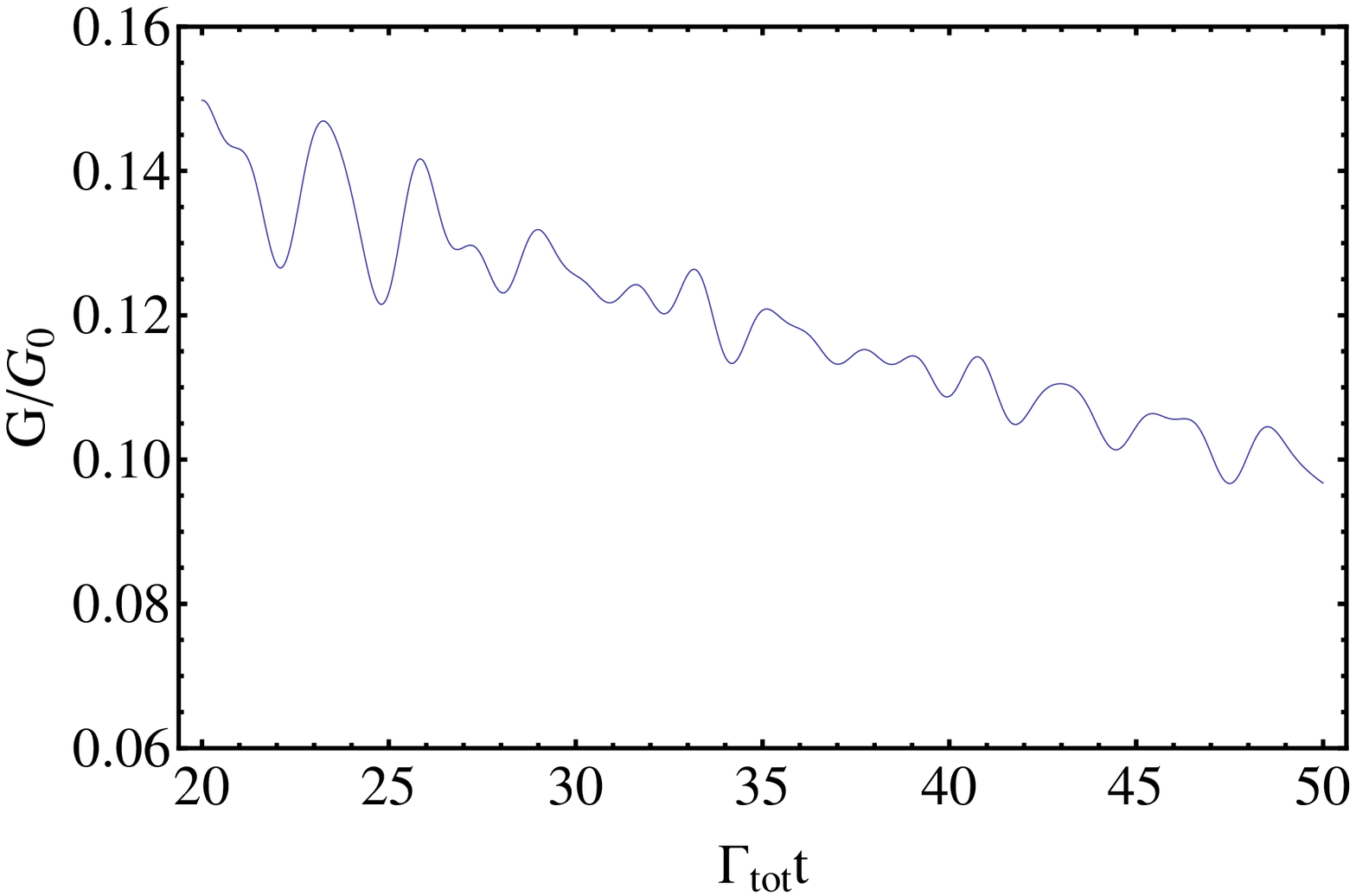} \\
\includegraphics[angle=0,width=6.8cm,height=5.0cm]{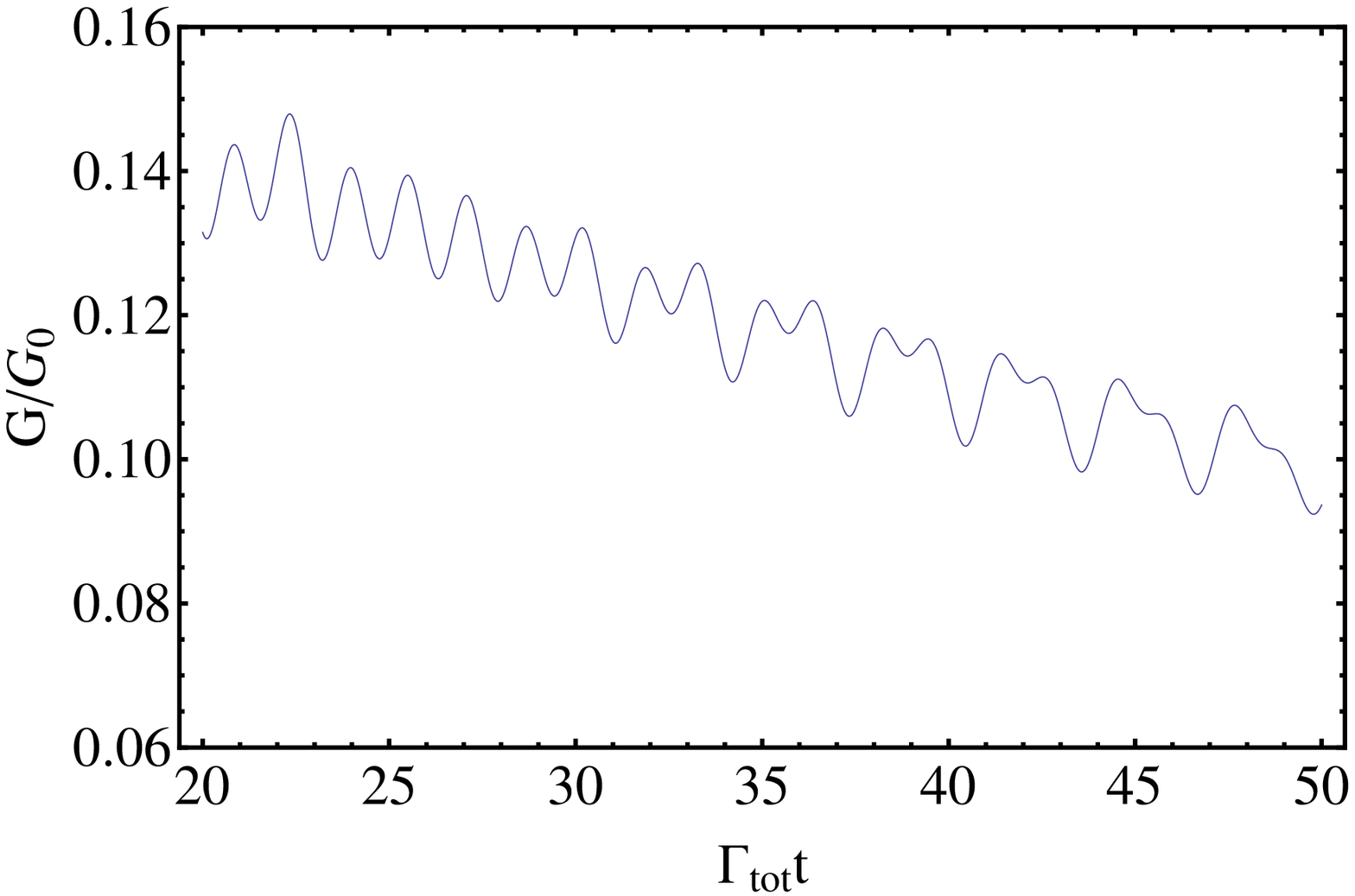} \\
\includegraphics[angle=0,width=6.8cm,height=5.0cm]{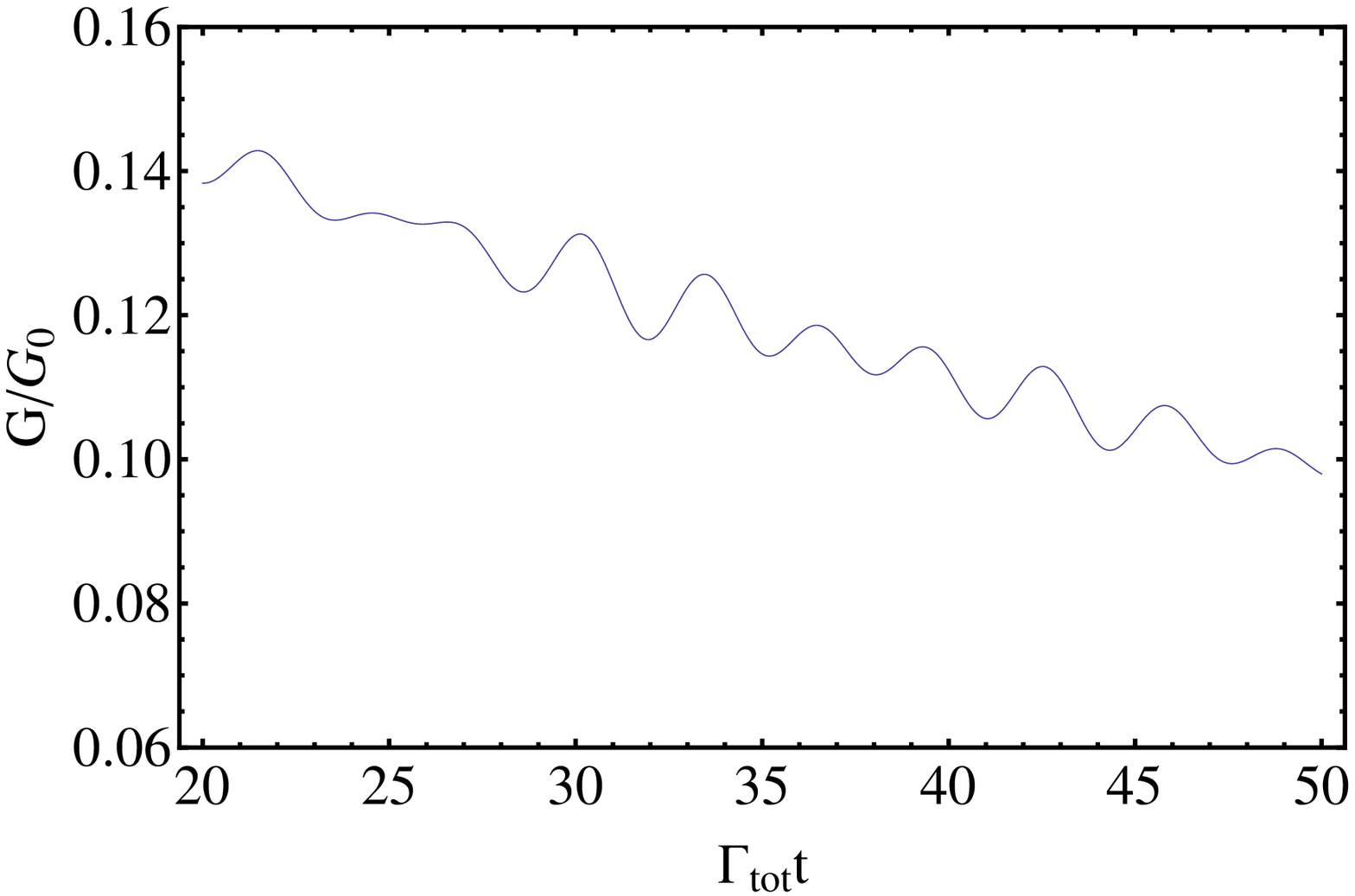}
\end{array}$
\end{center}
\caption{
Panels from top to bottom show the instantaneous conductance versus time in
Kondo timescale after the dot level has been switched to its final position for
(001) surface, (111) surface and (111) pyramid respectively with an asymmetry
factor of 0.9 at T=0.009$\Gamma_{tot}$ in infinitesimal bias.
}
\label{CPL3}
\end{figure}

Time dependent conductance results are displayed in 
Fig.~\ref{CPL3} for all three geometries in infinitesimal 
bias. Short timescale is omitted here due to the resemblance
to previous studies \cite{GokeretAl07JPCM}. The most 
striking feature in this figure is the extreme variations 
in conductance fluctuations among the three geometries. 
This effect can only originate from the difference in 
band structure of the contacts because all other 
parameters are kept fixed. We should point out that 
the slight difference between the Kondo temperature 
of (111) pyramide and the other two geometries does 
not change the relative behaviour of the fluctuations 
for these geometries because $T/T_K$ scaling has been 
shown to alter only the amplitude of the fluctuations 
\cite{GokeretAl07JPCM}. Consequently, the overall 
pattern is unaffected.

These fluctuations are obviously the result
of an admixture of sinusoidal oscillations with
different frequencies and amplitudes. Fourier
transform of the instantaneous conductance gives
each frequency component. It turns out that the 
frequencies are proportional to the separation 
between the peak positions and the Fermi level 
for all cases. This is precisely why instantaneous 
current for (001) surface exhibits a more erratic 
pattern compared to the others as five distinct 
frequencies contribute to it. Moreover, there are 
other peaks that appear in actual DFT data but 
they were excluded from our fitting since they 
have negligible contribution to the fluctuation 
pattern. The oscillation amplitude associated with 
them is quite small to cause a noticable effect 
because they are either located far away from 
the Fermi level or the peaks are not prominent 
compared to the surrounding features.

A microscopic mechanism for these instantaneous 
conductance results can be inferred with the
aid of the evolution of the dot density of states.
Based on the fact that discontinuities in the 
density of states of the contacts can interfere 
with the Kondo resonance when $\eta \ne 1$ \cite{GokeretAl07JPCM},
we propose that the fluctuations in time dependent 
conductance stem from the interference between the 
emerging Kondo resonance at the Fermi level and the 
sharp features in the density of states. Despite
the static nature of the sharp features, the Kondo 
resonance evolves in time in a dynamical way.
Consequently, the fluctuations continue until the 
Kondo resonance is fully formed. Furthermore, the 
reduction of oscillation amplitudes with increasing 
source-drain bias is related to the destruction of 
the Kondo resonance which diminishes the interference. 
This scenario is also supported by the saturation of 
oscillation amplitudes below the Kondo temperature 
because the Kondo resonance is fully developed below 
this energy scale. Consequently, the interference 
strength does not change upon lowering the ambient 
temperature any further.

\section{Vibrating SET}
\label{sec:vibrating}

In this section, we will present results
about a quantum dot which experiences
strong electron-phonon interaction. In our
approach, electron-phonon coupling term is
eliminated from the Holstein Hamiltonian
via Lang-Firsov canonical transformation.
This results in renormalized dot energy level
and Hubbard interaction strength. We then 
perform slave boson transformation and
invoke the non-crossing approximation
to solve the resulting Dyson equations.
The technical details of this implementation
can be found elsewhere \cite{Goker11JPCM}.

\begin{figure}[htbp]
\centerline{\includegraphics[angle=0,width=8.4cm,height=6.0cm]{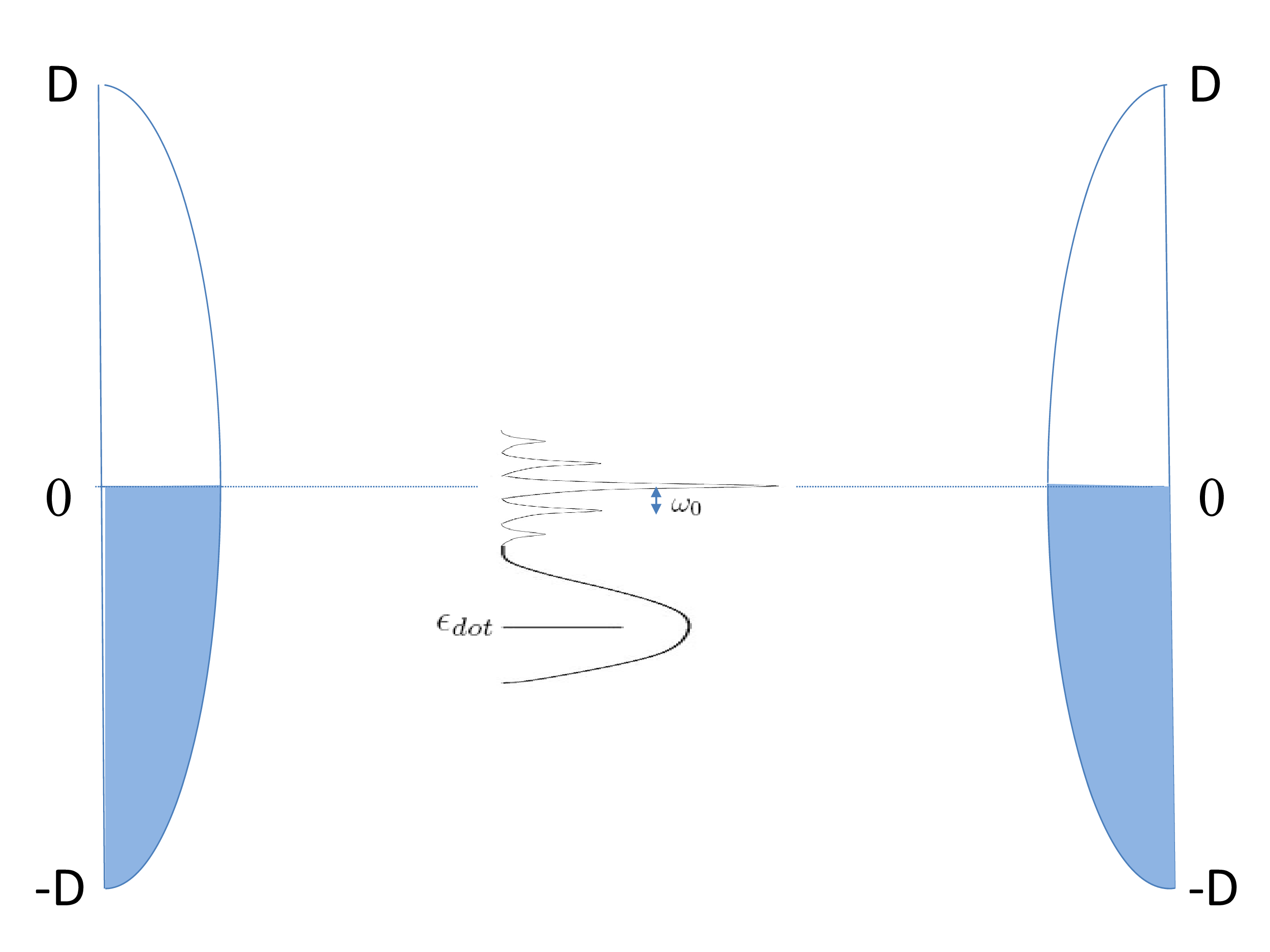}}
\caption{
This figure illustrates the density of states of both
contacts alongside with the application of an ac gate
voltage to the dot energy level as well as the
temperature gradient between the contacts.
}
\label{JPCM1}
\end{figure}

We will again consider the ramifications of
the sudden motion of the dot level from a 
position well below the Fermi level to a position 
where the Kondo effect is present with a gate 
voltage. Previous studies proved that the Kondo 
resonance which is pinned to the Fermi level
of the contacts acquires sidebands on each side
due to the electron-phonon coupling. Nevertheless, 
the transient dynamics of this system as a 
response to abrupt perturbations is unexplored. 
Fig.~\ref{JPCM1} depicts the system schematically.

\begin{figure}[htbp]
\centerline{\includegraphics[angle=0,width=8.4cm,height=6.0cm]{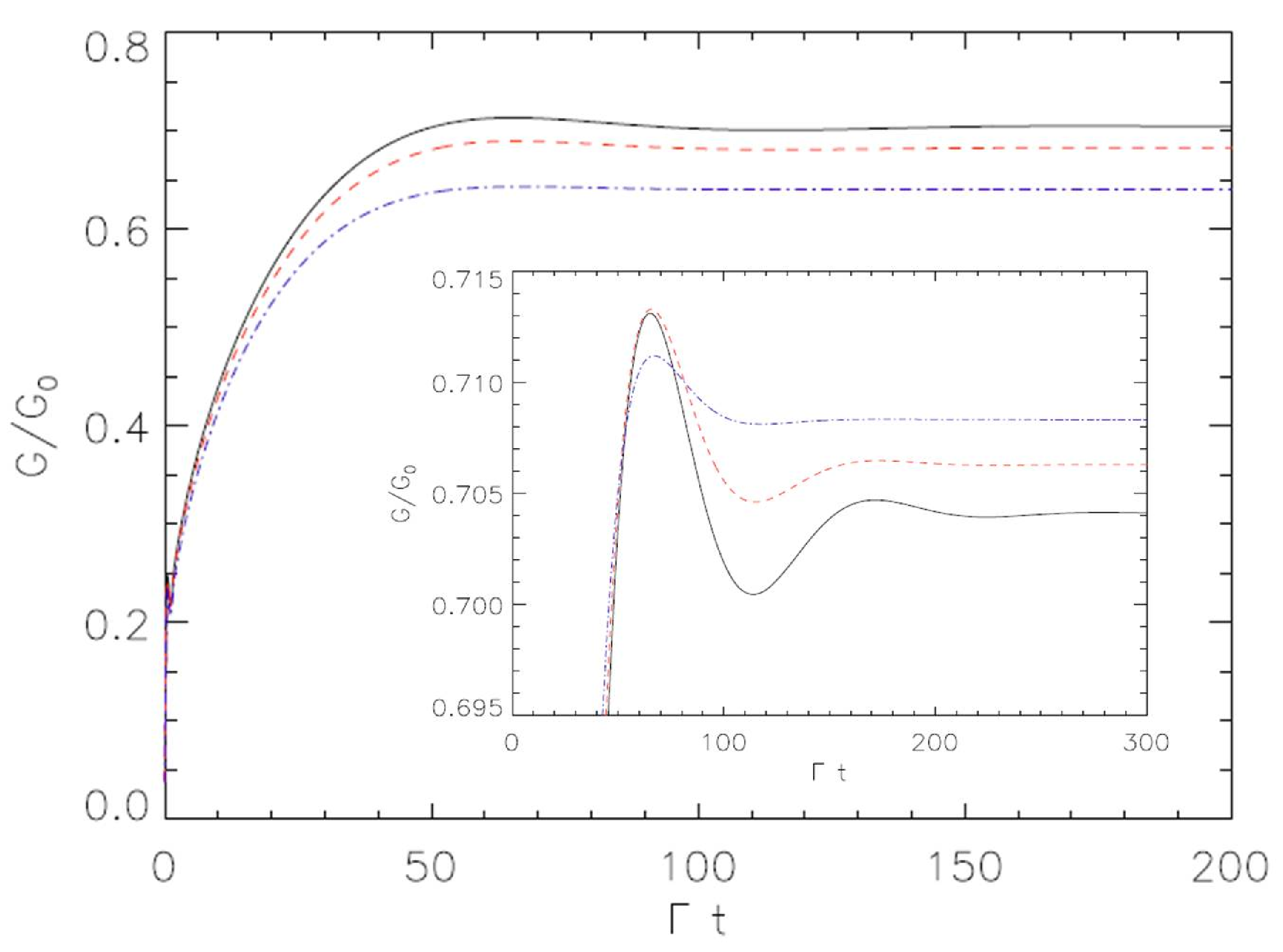}}
\caption{
This figure illustrates the density of states of both
contacts alongside with the application of an ac gate
voltage to the dot energy level as well as the
temperature gradient between the contacts.
}
\label{JPCM2}
\end{figure}

The instantaneous conductance results immediately 
after the dot level has been switched to its final 
position are shown in Fig.~\ref{JPCM2}. These data
have been taken for three different temperatures 
for a constant nonzero electron-phonon coupling $g$, 
which is defined as the square of the ratio of 
electron-phonon coupling strength $\lambda$ 
to phonon frequency $\omega$. Instantaneous 
conductance results for $g$=0 has been known 
for sometime \cite{PlihaletAl05PRB}. The Kondo 
timescale where the development of the Kondo 
resonance takes place occurs between 
10 $< \Gamma$ t $<$ 60 in Fig.~\ref{JPCM2}.

The steady state conductances (i.e. $ \Gamma t \rightarrow \infty$ ) 
are smaller than the case for $g$=0 in Fig.~\ref{JPCM2} 
for all temperatures. This is simply because the
electron-phonon coupling gradually suppresses the
Kondo resonance \cite{GalperinetAl07PRB}. This
suppression has been attributed to the downward
shift of the energy level due to the  phonon
reorganization. This originates from the 
renormalization of the dot level.

Another subtlety is the sinusoidal oscillations 
of the instantaneous conductance in the Kondo timescale.
These oscillations cannot be discerned in the main panel of
Fig.~\ref{JPCM2}. Main panel is magnified in the inset of 
Fig.~\ref{JPCM2} where conductance curves are shifted
in such a way that they overlap at the onset of oscillations.
It is obvious from here that the oscillation frequency 
is the same for all temperatures. Furthermore, the 
amplitude of the oscillations decreases with increasing 
ambient temperature and the oscillation frequency happens
to be equal to the phonon frequency $\omega_0$.

\begin{figure}[htbp]
\centerline{\includegraphics[angle=0,width=8.4cm,height=6.0cm]{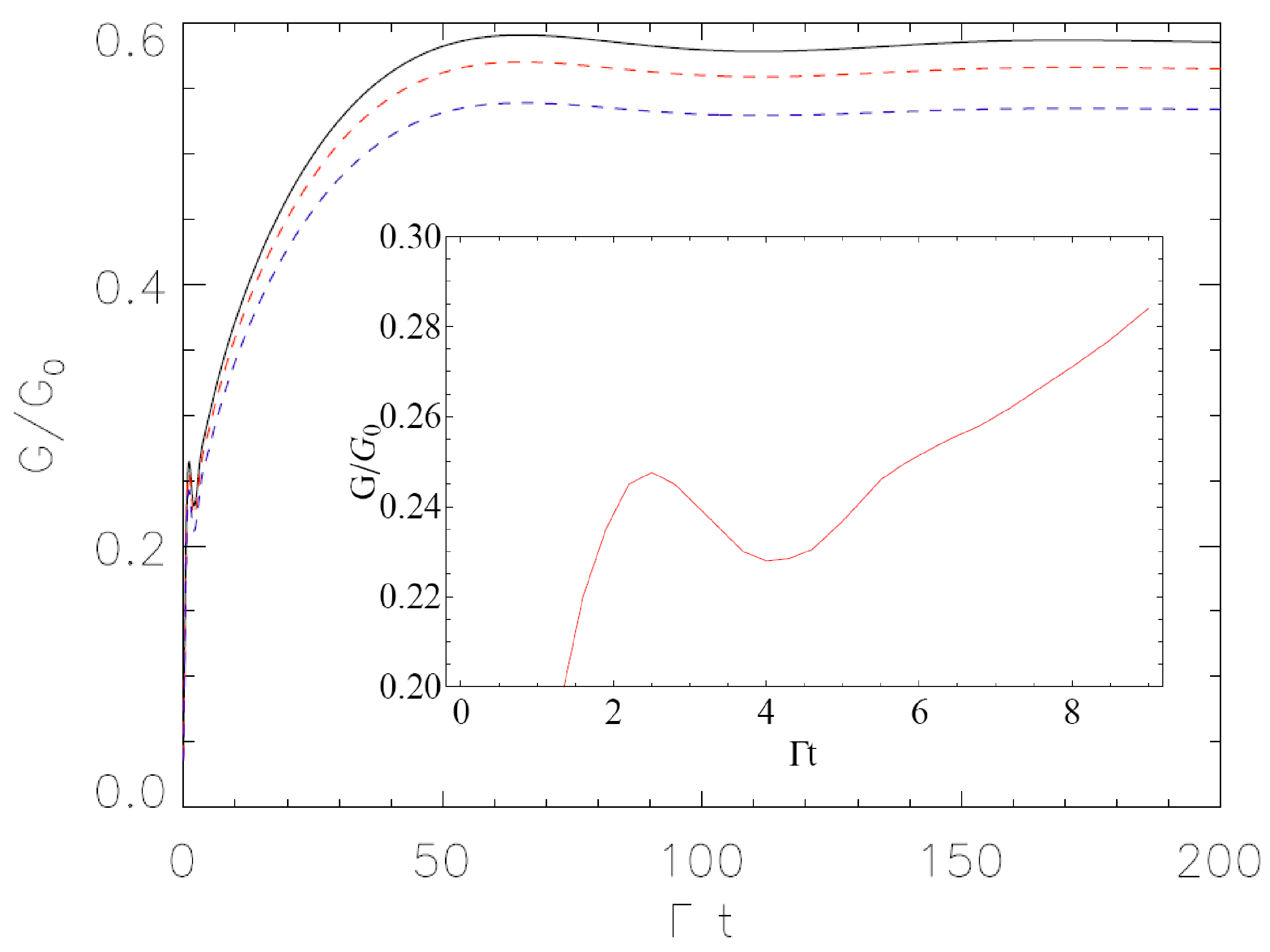}}
\caption{   
This figure illustrates the density of states of both
contacts alongside with the application of an ac gate
voltage to the dot energy level as well as the
temperature gradient between the contacts.
}
\label{JPCM3}
\end{figure}

We then performed the previous calculation at 
same ambient temperatures for a higher $g$ value 
to check whether oscillation frequency changes. 
Fig.~\ref{JPCM3} depicts the outcome. Increasing 
$g$ lowers the steady state conductance for all 
temperatures as anticipated because the the dot 
level is shifted down due to renormalization. More 
interestingly, frequency of the conductance oscillations 
is unchanged and again equal to the phonon
frequency $\omega_0$. This is a quite remarkable
and arresting result. We want to point out
that we changed $g$ in our calculations by increasing
$\lambda$ and keeping $\omega_0$ constant to 
enable a direct comparison. Further calculations
with other $g$ values proved that the oscillation 
frequency stays at $\omega_0$ for all $g$ values.

The instantaneous conductance exhibits oscillations
in the short timescale (i.e. 0 $< \Gamma$ t $<$ 10)
related to the charge transfer \cite{PlihaletAl05PRB,MuhlbacheretAl08PRL}
as well. These oscillations are shown in the
inset of Fig.~\ref{JPCM3}. These oscillations
are not related to the formation of the Kondo
resonance because spin-flip processes are 
yet to begin here. The frequency of these
oscillations have been found to be proportional 
to the final dot level \cite{PlihaletAl05PRB}.

\begin{figure}[htbp]
\centerline{\includegraphics[angle=90,width=9.0cm,height=6.7cm]{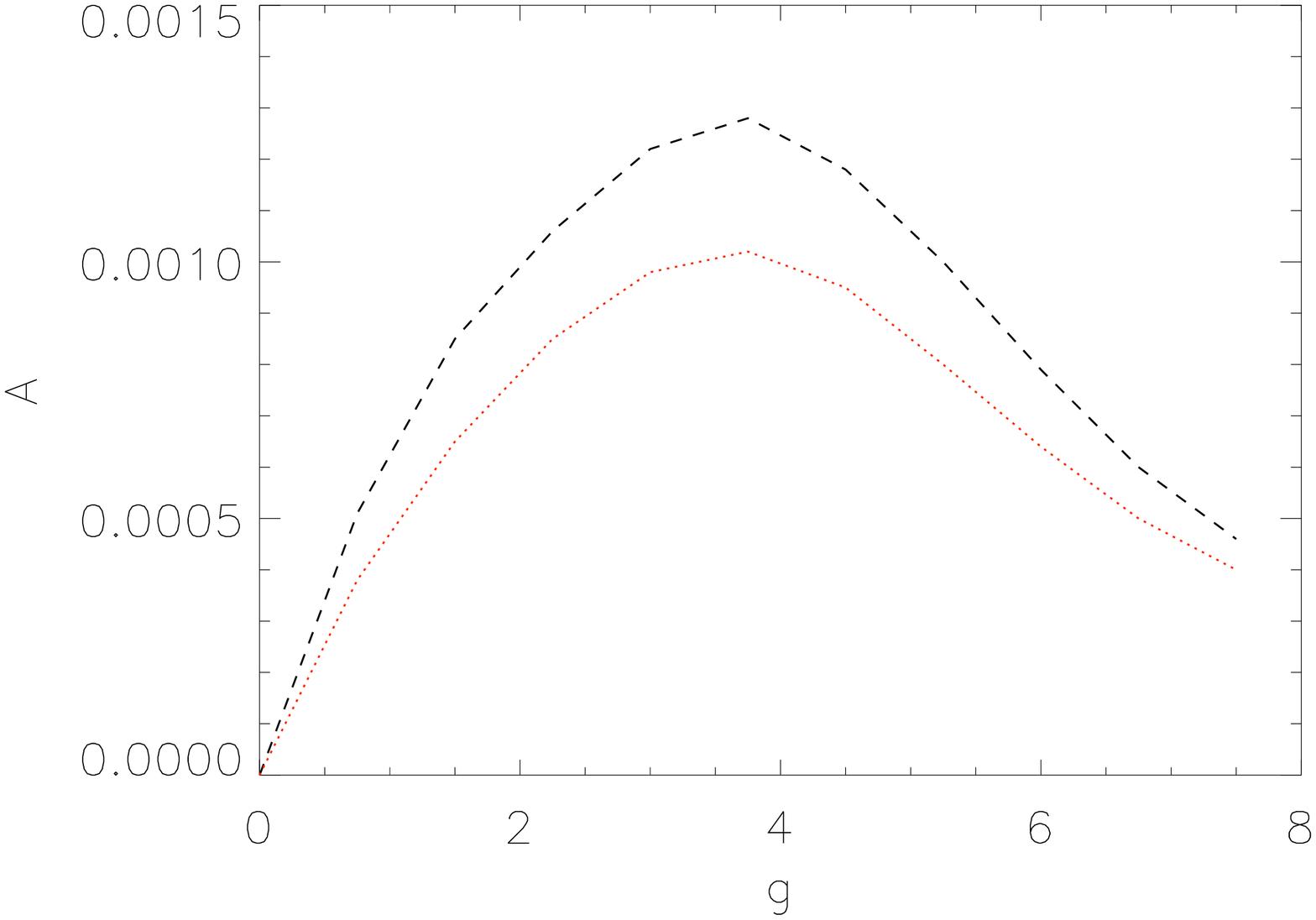}}
\caption{   
This figure illustrates the density of states of both
contacts alongside with the application of an ac gate
voltage to the dot energy level as well as the
temperature gradient between the contacts.
}
\label{JPCM4}
\end{figure}

Fig.~\ref{JPCM4} displays the behaviour of the 
amplitude of oscillations as a function of $g$ at 
two different temperatures. The amplitude is zero
at $g$=0 for both cases but it starts to increase 
gradually until it reaches a peak before $g$=4. It 
then starts decreasing and approaches zero for large 
$g$ values. Furthermore, we see that lowering temperature
results in larger oscillation amplitudes for all 
$g$ values. Our calculations at other temperatures 
showed that this conclusion is general and valid 
for other ambient temperatures too.

We propose the interference between the 
main Kondo resonance and its phonon sidebands
as the underlying microspopic mechanism for 
the sinusoidal oscillations seen in the 
instantaneous conductance in the Kondo 
timescale. Previous investigations demonstrated 
clearly that the phonon sidebands are separated 
from the main Kondo peak by an integer multiple 
of phonon frequency $\omega_0$
\cite{ChenetAl05PRB,GalperinetAl06PRB,WangetAl07PRB,PaaskeetAl05PRL}.
That is why the frequency of oscillations
always turns out to be equal to $\omega_0$.
For a given $g$, the amplitude of oscillations 
increases with decreasing ambient temperature 
because both the main Kondo peak and its phonon 
satellites are more robust. Consequently, 
interference between them is stronger. When 
$g$=0, the amplitude of oscillations is zero 
because phonon sidebands do not exist in this 
limit. Even though the main Kondo peak is 
quenched somewhat for small $g$, its phonon
sidebands become slightly more pronounced
\cite{YangetAl10EPL}, thus the amplitude 
of oscillations starts to grow. Nevertheless, 
the amplitudes start decreasing around $g$=4 
since all peaks start to get inhibited. The 
amplitude vanishes for large $g$ where the 
Kondo effect is destroyed completely and all 
peaks are gone.

\section{Thermoelectric effects}
\label{sec:thermal}

In this chapter, we will be interested in Seebeck 
coefficient or thermopower of this device. In linear
response, it is given by
\begin{equation}
S(t)=\frac{L_{12}(t)}{T L_{11} (t)},
\label{Seebeck}
\end{equation}
where the Onsager coefficients are
\begin{eqnarray}
& & L_{11} (t)= T \times \nonumber \\
& & Im \left(\int_{-\infty} ^t dt_1 \int \frac{d\epsilon}{2\pi} e^{i\epsilon(t-t_1)} \Gamma(\epsilon) G^r (t,t_1) \frac{\partial f(\epsilon)} {\partial \epsilon}\right)
\label{Onsager1}
\end{eqnarray}
and
\begin{eqnarray}
& &L_{12} (t)= T^2 \times \nonumber \\
& & Im \left(\int_{-\infty} ^t dt_1 \int \frac{d\epsilon}{2\pi} e^{i\epsilon(t-t_1)} \Gamma(\epsilon) G^r (t,t_1) \frac{\partial f(\epsilon)} {\partial T}\right).
\label{Onsager2}
\end{eqnarray}

\begin{figure}[htbp]
\centerline{\includegraphics[angle=0,width=9.2cm,height=6.6cm]{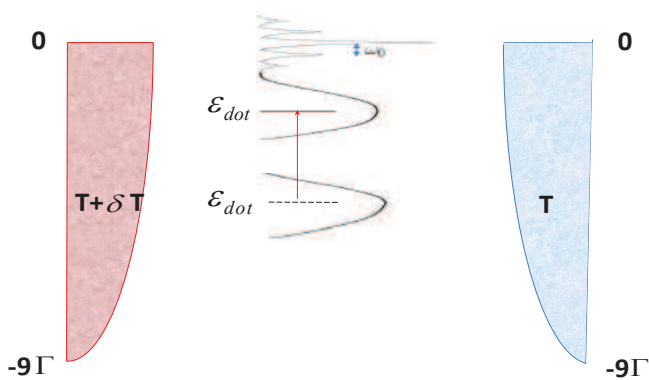}}
\caption{
This figure shows the density of states of both contacts and
the quantum dot in the initial and final states schematically.
The temperature gradient between the contacts is also depicted.
}
\label{PLA1}
\end{figure}

We will first study the behaviour of the 
instantaneous thermopower immediately after 
the dot level is switched from $\epsilon_1=-5\Gamma$ 
to $\epsilon_2=-2\Gamma$ at t=0 via a gate voltage 
for a symmetrically coupled system. Kondo resonance 
starts to emerge in the final state as a result of
this transition. Fig.\ref{PLA1} shows the density of 
states of the dot both in initial and final levels 
plus the the density of states of the contacts
schematically.

\begin{figure}[htbp]
\centerline{\includegraphics[angle=0,width=8.4cm,height=6.0cm]{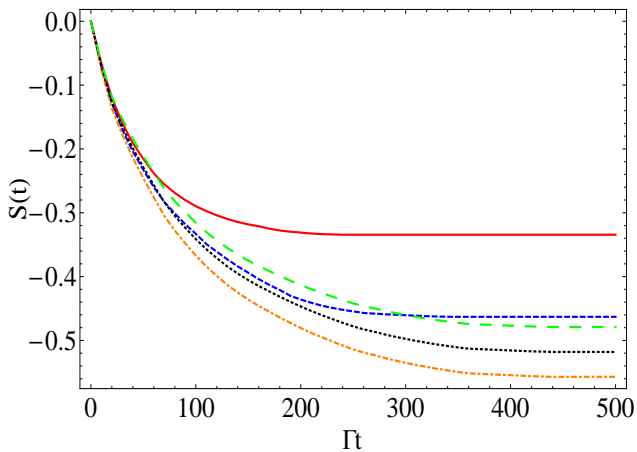}}
\caption{
This figure shows the instantaneous thermopower $S(t)$
immediately after the dot level has been moved to its
final position for $T_1$=0.0035$\Gamma$ (red solid),
$T_2$=0.0028$\Gamma$ (blue short dashed),
$T_3$=0.0021$\Gamma$ (orange dot dashed),
$T_4$=0.0014$\Gamma$ (black dotted),
$T_5$=0.0007$\Gamma$ (green long dashed)
in linear response without any electron-phonon coupling.
}
\label{PLA2}
\end{figure}

Fig. \ref{PLA2} shows instantaneous thermopower
at various ambient temperatures immediately after 
the dot level was switched to its final position
in infinitesimal bias for $g$=0. It is clear that
the transient thermopower decays from zero to its
steady state value for all temperatures and
this steady state value is always negative. 
On the other hand, the absolute value of steady
state thermopower goes up with decreasing
temperature until it reaches a maximum at a 
critical temperature. It starts to decrease if the
temperature is lowered any further. However,
the decay time, defined as the time required
to reach the steady state value, remains constant.
This critical temperature has been shown to be
equal to the Kondo temperature \cite{CostietAl10PRB}.
Saturation of the decay time of thermopower below
Kondo temperature is the new physics our results 
reveal.

\begin{figure}[htbp]
\centerline{\includegraphics[angle=0,width=8.4cm,height=6.0cm]{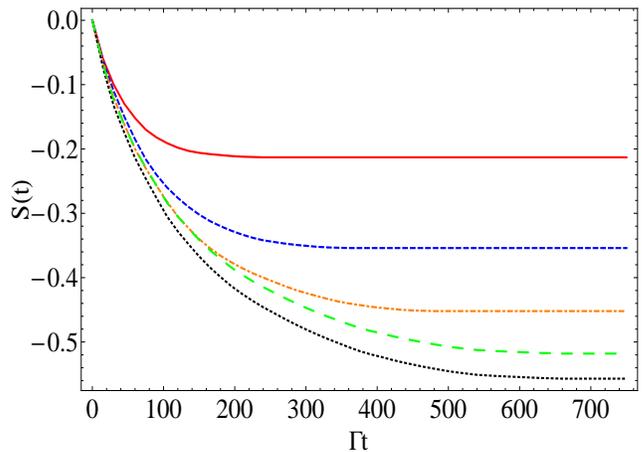}}
\caption{
This figure shows the instantaneous thermopower $S(t)$
immediately after the dot level has been moved to its
final position for $T_1$=0.0035$\Gamma$ (red solid),
$T_2$=0.0028$\Gamma$ (blue short dashed),
$T_3$=0.0021$\Gamma$ (orange dot dashed),
$T_4$=0.0014$\Gamma$ (black dotted),
$T_5$=0.0007$\Gamma$ (green long dashed)
in linear response for $g$=2.25.
}
\label{PLA3}
\end{figure}

Fig.~\ref{PLA3} displays the instantaneous thermopower
for the same parameters used in Fig.~\ref{PLA2}. The 
difference is that we turn on electron-phonon interaction
and take $g$=2.25 with $\omega_0$=0.08$\Gamma$. In this case,
the dot level is shifted slightly downwards due to
renormalization leading to a smaller Kondo temperature
Smaller Kondo temperature due to the renormalization
of the dot level cause steady state thermopower
to be less than $g$=0 case provided that the ambient
temperature is above both Kondo temperatures. However, 
steady state thermopower exceeds the value attained 
at $g$=0 case when the ambient temperature falls 
below both Kondo temperatures.

The major difference with $g$=0 case is the duration
of the saturated decay time. It is now much longer 
as lowest curves in both Fig.~\ref{PLA2} and Fig.~\ref{PLA3} 
clearly show. The decay time of thermopower scales with
$1/T_K$ according to our calculations. This scaling is
reminiscent to previous time dependent conductance calculations
\cite{PlihaletAl05PRB}. The fundamental mechanism
is the same for both situations because they rely on
the evolution of Kondo resonance whose development is 
inversely proportional to the Kondo temperature 
\cite{NordlanderetAl99PRL}.

We want to provide a microscopic explanation
for the numerical results presented above.
The relation between the dot density of states 
and the thermopower at low temperatures is 
given by the Sommerfeld expansion. It can be 
cast as
\begin{equation}
S(T)=-\frac{\pi^2 T}{3 A(0,T)}\frac{\partial A}{\partial \epsilon} \rvert_{\epsilon=0}
\end{equation}
in atomic units. Here, $A(0,T)$ is the value of 
the dot density of states at Fermi level
and $\frac{\partial A}{\partial \epsilon}$
is its derivative. The Kondo resonance is formed
slightly above the Fermi level \cite{CostietAl94JPCM} 
causing the derivative of the dot density of states 
at Fermi level to be positive. This will naturally 
generate a negative thermopower in Kondo regime. 
Otherwise, the thermopower will be positive without 
the Kondo resonance because the only prominent 
feature of the dot density of states is the 
Breit-Wigner resonance in that case.

The Kondo resonance is in its most developed form 
when T$\le T_K$. Therefore, the decay time of thermopower
stays constant below $T_K$. On the other hand, 
the steady state value keeps decreasing due to the 
$T$ prefactor at the beginning of Sommerfeld expansion 
which is the only variable below $T_K$.
Furthermore, $S(t)$=0 at $t$=0 because the Kondo
resonance does not exist when the dot is
in its initial level since T$\gg T_K$. This means
that the dot density of states is essentially flat 
at Fermi level. This in turn implies zero slope 
resulting in zero thermopower.

Finally, we want to investigate the behaviour of
Eq.~(\ref{Seebeck}) for a quantum dot in Kondo
regime whose energy level is displaced sinusoidally 
via a gate voltage. Time averaged thermopower has 
been previously investigated for this set-up without 
the Kondo resonance  \cite{ChietAl12JPCM}. 
It is quite intriguing to add strong correlation
effects to this model since they play a significant
role in confined nanostructures like quantum dots.
We will again be restricted to the linear response 
thermopower because the Onsager relations are valid 
only in this regime.

\begin{figure}[htb]
\centerline{\includegraphics[angle=0,width=9.2cm,height=6.6cm]{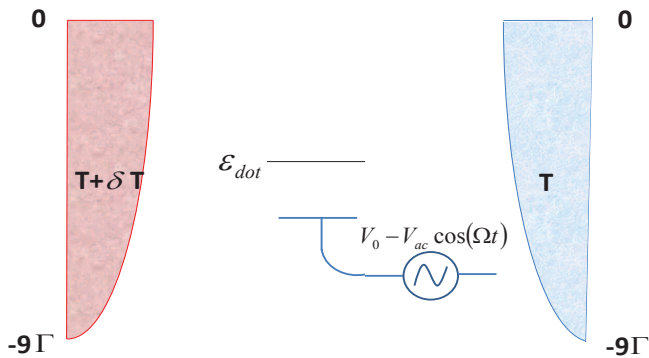}}
\caption{
This figure illustrates the density of states of both
contacts alongside with the application of an ac gate
voltage to the dot energy level as well as the
temperature gradient between the contacts.
}
\label{PRL1}
\end{figure}

We can represent the instantaneous behaviour 
of the dot energy level as
\begin{equation}
\epsilon_{dot}(t)=-5\Gamma-A cos(\Omega t).
\label{dotlevel}
\end{equation}
where $A$ is the driving amplitude and $\Omega$ 
is the driving frequency. This kind of perturbation 
causes a continuous implicit change in Kondo temperature 
$T_K$ because the dot energy level affects $T_K$ 
according to Eq.~(\ref{tkondo}). This results in a 
constant change in the shape of the Kondo resonance. 
The situation under consideration is depicted 
schematically in Fig.~\ref{PRL1}.

\begin{figure}[htb]
\centerline{\includegraphics[angle=0,width=8.5cm,height=5.8cm]{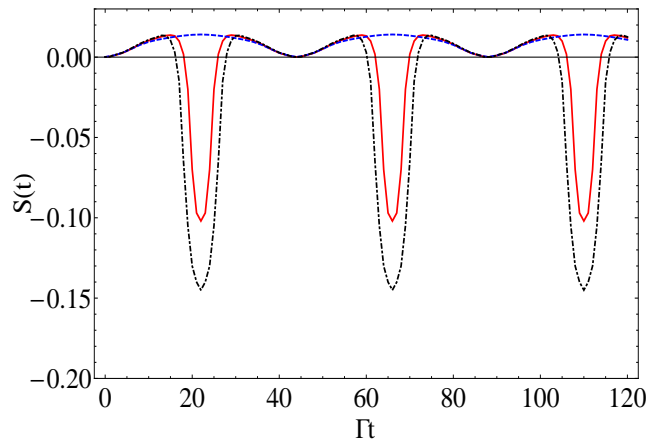}}
\caption{
This figure shows the instantaneous thermopower
$S(t)$ immediately after the gate voltage has
been turned on for driving amplitudes of
$2.0\Gamma$ (blue dashed),
$2.5\Gamma$ (red solid) and
$3.0\Gamma$ (black dot dashed)
at $T$=0.003$\Gamma$ and
$\Omega$=0.14$\Gamma$
in linear response.
}
\label{PRL2}
\end{figure}

Fig.~\ref{PRL2} shows instantaneous thermopower
for three different driving amplitudes $A$ at fixed
driving frequency and ambient temperature after
the gate voltage is applied. Thermopower 
slowly starts to increase as the dot level 
approaches the Fermi level for all driving 
amplitudes. For the smallest driving amplitude, 
it reaches a peak value at its closest point to 
the Fermi level and then gradually decreases
again to zero as the dot level is moved away
from the Fermi level. This monotonic behaviour 
is altered as the driving amplitude is ramped up.
When $A=2.5\Gamma$, the instantaneous thermopower 
initially grows in the positive region similar to 
the previous case. This similarity ends as the dot 
level comes close to the Fermi level because the
instantaneous thermopower abruptly starts to decrease
and dives into negative region before bouncing
back. In ever higher driving amplitudes, instantaneous 
thermopower starts its descent earlier. Furthermore,
it goes deeper into negative territory before
rebounding with retreating dot level. These
fluctuations are repeated in every period
of dot level oscillation.

\begin{figure}[htb]
\centerline{\includegraphics[angle=0,width=8.4cm,height=5.8cm]{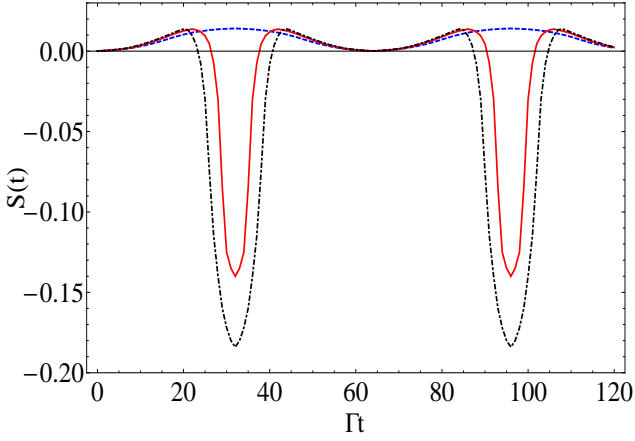}}
\caption{
This figure shows the instantaneous thermopower
$S(t)$ immediately after the gate voltage has
been turned on for driving amplitudes of
$2.0\Gamma$ (blue dashed),
$2.5\Gamma$ (red solid) and
$3.0\Gamma$ (black dot dashed)
at $T$=0.003$\Gamma$ and
$\Omega$=0.10$\Gamma$  
in linear response.
}
\label{PRL3}
\end{figure}

In Fig.~\ref{PRL3}, we calculate the instantaneous 
thermopower with a smaller driving frequency than 
the one in Fig.~\ref{PRL2}. All other parameters
are unchanged. This choice naturally leads to longer 
oscillation periods for all driving amplitudes. 
Beyond this triviality, we see that the fairly 
smooth oscillation pattern for the smallest driving 
amplitude is still intact. The fluctuations of the 
other two larger driving amplitude cases also look
similar to the larger driving frequency case but 
they both dive much deeper into negative territory 
when the dot level is in the vicinity of the Fermi level.

\begin{figure}[htb]
\centerline{\includegraphics[angle=0,width=8.4cm,height=5.8cm]{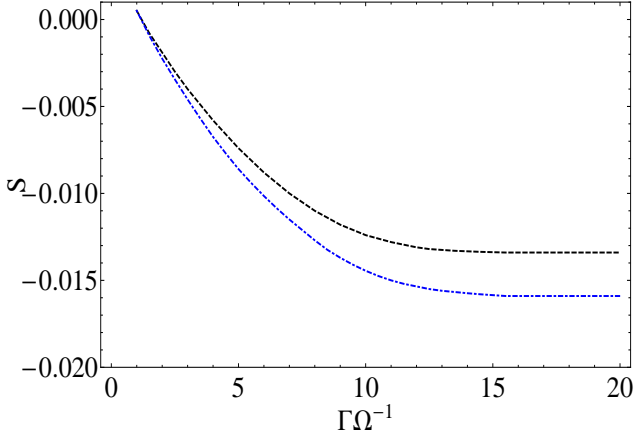}}
\caption{
This figure shows the value of the 
thermopower averaged over a period as a function
of inverse driving frequency at $T$=0.003$\Gamma$
(black dashed) and $T$=0.002$\Gamma$ (blue dot dashed)
for driving amplitude of 2.5$\Gamma$.
}
\label{PRL4}
\end{figure}

In order to determine whether these particular
outcomes are valid for other driving frequencies 
and amplitudes, we performed an analysis by 
averaging the value of the instantaneous thermopower
over a full period for two different ambient 
temperatures at several different driving
frequencies. We will omit the smallest driving 
amplitude because time averaged thermopower 
was insensitive to the changes in temperature for 
it. The time averaged value of the instantaneous 
thermopower is shown in Fig.~\ref{PRL4} as a function
of inverse driving frequency for two different
ambient temperatures with $A=2.5\Gamma$. First 
of all, altering the ambient temperature
results in a change in absolute value of 
the time averaged thermopower at all driving 
frequencies. Furthermore, absolute value of 
the time averaged thermopower starts to increase 
at both temperatures with decreasing driving
frequency but it saturates below a certain 
driving frequency.

\begin{figure}[htb]
\centerline{\includegraphics[angle=0,width=8.2cm,height=5.6cm]{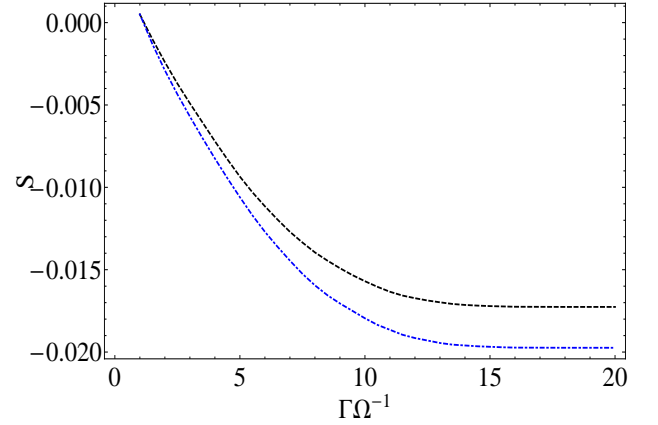}}
\caption{
This figure shows the value of the
thermopower averaged over a period as a function
of inverse driving frequency at $T$=0.003$\Gamma$
(black dashed) and $T$=0.002$\Gamma$ (blue dot dashed)
for driving amplitude of 3.0$\Gamma$.
}
\label{PRL5}
\end{figure}

We then extended this analysis to the 
largest driving amplitude of 3.0$\Gamma$ 
keeping all other parameters constant. The
result is seen in Fig.~\ref{PRL5} which
confirms the previous conclusions about
the sensitivity of the time averaged thermopower
to ambient temperature and its saturation 
below a certain driving frequency. However, 
absolute values of the time averaged thermopower 
at both ambient temperatures and all driving 
frequencies are greater than the case in 
Fig.~\ref{PRL4} with 2.5$\Gamma$.

The dot level cannot sufficiently come close to 
the Fermi level to acquire the Kondo resonance 
when the driving amplitude is 2.0$\Gamma$.
This means that T$\gg T_K$ throughout this oscillation 
and consequently the thermopower manages to stay in 
positive territory. It is zero at the furthest position 
of the dot level from the Fermi level since the 
density of states vanishes around the Fermi level. 
It then slowly starts to increase as the dot
level approaches the Fermi level because the tail 
of the Breit-Wigner resonance starts to touch the 
Fermi level giving rise to a negative slope. Thermopower
hits a peak positive value when the dot level is
closest to the Fermi level. Thermopower starts to
decrease as the dot level starts to move away.
The lack of sensitivity to ambient temperature 
for this driving amplitude is related to the
absence of Kondo resonance everywhere in this
motion.

Boosting the driving amplitude to 2.5$\Gamma$ makes
the oscillation of the dot energy level cover a larger
energy range. The instantaneous thermopower initially 
follows the curve obtained for 2.0$\Gamma$. 
Nevertheless, it begins to deviate considerably 
when the dot level approaches the Fermi level due to 
the emergence of the Kondo resonance. This implies
a shift in slope of the density of states of the dot 
accompanied by a change in sign of the thermopower. 
When the dot level is at its closest point to the 
Fermi level, thermopower reaches a bottom since the 
Kondo resonance is in its most developed shape at that 
point. When the dot level starts to move away from the 
Fermi level, the Kondo resonance disappears gradually, 
hence the thermopower climbs out of the bottom becoming
positive after a while. The instantaneous thermopower 
curve is quite similar for the largest driving amplitude 
of 3.0$\Gamma$ except two differences. The instantaneous 
thermopower dips lower and earlier into negative region 
simply because the dot energy level is closer to the 
Fermi level at any given time.

The evolution of the Kondo resonance is
also the underlying factor behind  
the saturation of the time averaged 
thermopower at low driving frequencies. 
As the driving frequency is lowered,
The dot energy level finds the chance to 
spend more time near the Fermi level with 
decreasing driving frequency. Since the 
formation of the Kondo resonance requires 
long time \cite{NordlanderetAl99PRL}, this
sets aside more time for the Kondo resonance
to develop. As a result, time averaged 
thermopower grows in negative territory 
until it saturates below a certain driving 
frequency. Saturation stems from the 
allocation of sufficient time for the
full development of the Kondo resonance.

Finally, we want to elaborate on the 
dependence of the time averaged thermopower
on temperature. Both Fig.~\ref{PRL4} and 
Fig.~\ref{PRL5} reveal that the difference 
between the time averaged values at the 
same driving frequency increases with 
decreasing driving frequency before
saturating below a certain frequency.
This underpins the significance of the 
development of the Kondo resonance. 
The lack of sensitivity of time averaged
thermopower to ambient temperature at 
elevated driving frequencies stems from 
the insufficient time allocated to the dot
level to spend around the Fermi level. This 
prevents the development of the Kondo resonance 
to its full extent. As the driving frequency 
is reduced, the difference becomes more discernible 
with decreasing driving frequency until it
saturates at some point because the Kondo 
resonance is fully developed below that frequency.

\section{Conclusion}
 
In this chapter, we summarized the recent developments
in time dependent electron transport through a 
strongly correlated single electron transistor.
In section~\ref{sec:designer}, we demonstrated 
that the discontinuities in the density of 
states of the electrodes can induce oscillations 
in the transient current whose frequency is
proportional to the energetic separation between
the Fermi level and the position of the discontinuity.
Strong electron-phonon coupling on the dot 
has been shown in section ~\ref{sec:vibrating} 
to give rise to oscillations in transient current 
whose frequency is equal to the phonon frequency.
Finally, we investigated the evolution of Seebeck 
coefficient or thermopower in section ~\ref{sec:thermal}. 
For the case where the dot level is abruptly changed, 
we determined that the decay time of thermopower
to its steady state value saturates below the 
Kondo temperature. On the other hand, the 
instantaneous thermopower dips into negative
region if the driving amplitude is ramped up
such that the Kondo resonance finds time to
develop.

We think that our findings are crucial to 
design the switching behaviour of next generation 
transistors that are expected to replace the 
current MOSFET's in near future. In that sense,
our results have widespread ramifications
along several disciplines such as physics, 
material science and electrical engineering. 
We also believe that state of the art ultrafast pump-probe 
techniques \cite{Teradaetal10JPCM,TeradaetAl10Nature}
enable the experimental realization of our
results. Consequently, our results in this
chapter should pave the way for new experiments 
in this field.

\section{Acknowledgments}

Both authors thank T$\ddot{u}$bitak for
generous financial support via grant 111T303.
AG is indebted to Dr. Udo Schwingenschlogl
and Dr. Xuhui Wang for several fruitful
discussions and thanks Dr. Peter
Nordlander for introducing this subject 
to him many years ago.

\bibliographystyle{iopams}

\begin{thebibliography}{10}
\expandafter\ifx\csname url\endcsname\relax
  \def\url#1{{\tt #1}}\fi
\expandafter\ifx\csname urlprefix\endcsname\relax\def\urlprefix{URL }\fi
\providecommand{\eprint}[2][]{\url{#2}}

\bibitem{Likharev03}
Likharev K~K 2003 in J~Greer, A~Korkin and J~Labanowski, eds, {\em Nano and
  Giga challenges in microelectronics\/} (Dordrecht: Elsevier) pp 27--68

\bibitem{ElzermanetAl04Nature}
Elzerman J~M, Hanson R, van Beveren L~H~W, Witkamp B, Vandersypen L~M~K and
  Kouwenhoven L~P 2004 {\em Nature(London)\/} {\bf 430} 431--434

\bibitem{FeveetAl07Science}
Feve G, Mahe A, Berroir J~M, Kontos T, Placais B, Glattli D~C, Cavanna A,
  Etienne B and Jin Y 2007 {\em Science\/} {\bf 316} 1169

\bibitem{PlihaletAl05PRB}
Plihal M, Langreth D~C and Nordlander P 2005 {\em Phys. Rev. B\/} {\bf 71}
  165321

\bibitem{IzmaylovetAl06JPCM}
Izmaylov A~F, Goker A, Friedman B~A and Nordlander P 2006 {\em J. Phys.:
  Condens. Matter\/} {\bf 18} 8995--9006

\bibitem{NordlanderetAl99PRL}
Nordlander P, Pustilnik M, Meir Y, Wingreen N~S and Langreth D~C 1999 {\em
  Phys. Rev. Lett.\/} {\bf 83} 808--811

\bibitem{PlihaletAl00PRB}
Plihal M, Langreth D~C and Nordlander P 2000 {\em Phys. Rev. B\/} {\bf 61}
  R13341--13344

\bibitem{MerinoMarston04PRB}
Merino J and Marston J~B 2004 {\em Phys. Rev. B\/} {\bf 69} 115304

\bibitem{GokeretAl07JPCM}
Goker A, Friedman B~A and Nordlander P 2007 {\em J. Phys.: Condens. Matter\/}
  {\bf 19} 376206

\bibitem{GokeretAl10PRB}
Goker A, Zhu Z~Y, Manchon A and Schwingenschlogl U 2010 {\em Phys. Rev. B\/}
  {\bf 82} 161304(R)

\bibitem{GokeretAl11CPL}
Goker A, Zhu Z~Y, Manchon A and Schwingenschlogl U 2011 {\em Chem. Phys.
  Lett.\/} {\bf 509} 48

\bibitem{Goker11JPCM}
Goker A 2011 {\em J. Phys.:Condens. Matter\/} {\bf 23} 125302

\bibitem{Reddyetal07Science}
Reddy P, Jang S~Y, Segalman R~A and Majumdar A 2007 {\em Science\/} {\bf 315}
  1568

\bibitem{Bahetietal08NL}
Baheti K, Malen J~A, Doak P, Reddy P, Jang S~Y, Tilley T~D, Majumdar A and
  Segalman R~A 2008 {\em Nano Lett.\/} {\bf 8} 715

\bibitem{Malenetal09NL}
Malen J~A, Doak P, Baheti K, Tilley T~D, Majumdar A and Segalman R~A 2009 {\em
  Nano Lett.\/} {\bf 9} 3406

\bibitem{TangetAl10APL}
Tan A, Sadat S and Reddy P 2010 {\em Appl. Phys. Lett.\/} {\bf 96} 13110

\bibitem{DongetAl02JPCM}
Dong B and Lei X~L 2002 {\em J. Phys.:Condens. Matter\/} {\bf 14} 11747

\bibitem{CostietAl10PRB}
Costi T~A and Zlatic V 2010 {\em Phys. Rev. B\/} {\bf 81} 235127

\bibitem{Goker2012}
Goker A and Uyanik B 2012 {\em Phys. Lett. A\/} {\bf 376} 2735

\bibitem{AlhassaniehetAl05PRL}
Al-Hassanieh K~A, Busser C~A, Martins G~B and Dagotto E 2005 {\em Phys. Rev.
  Lett.\/} {\bf 95} 256807

\bibitem{KiselevetAl06PRB}
Kiselev M~N, Kikoin K, Shekhter R~I and Vinokur V~M 2006 {\em Phys. Rev. B\/}
  {\bf 74} 233403

\bibitem{ScheibleetAl04PRL}
Scheible D~V, Weiss C, Kotthaus J~P and Blick R~H 2004 {\em Phys. Rev. Lett.\/}
  {\bf 93} 186801

\bibitem{ParksetAl07PRL}
Parks J~J, Champagne A~R, Hutchison G~R, Flores-Torres S, Abruna H~D and Ralph
  D~C 2007 {\em Phys. Rev. Lett.\/} {\bf 99} 026601

\bibitem{Goker08SSC}
Goker A 2008 {\em Solid State Comm.\/} {\bf 148} 230

\bibitem{Gokeretal2012}
Goker A and Gedik E 2012 submitted

\bibitem{ShaoetAl194PRB}
Shao H~X, Langreth D~C and Nordlander P 1994 {\em Phys. Rev. B\/} {\bf 49}
  13929--13947

\bibitem{Blahaetal01Book}
Blaha P, Schwarz K, Madsen G~K~H, Kvasnicka D and Luitz L 2001 {\em WIEN2K, an
  augmented plane wave+local orbitals program for calculating crystal
  properties\/} (Wien: Techn. Universit$\ddot{a}$t)

\bibitem{Perdewetal96PRL}
Perdew J~P, Burke K and Ernzerhof M 1996 {\em Phys. Rev. Lett.\/} {\bf 77} 3865

\bibitem{JauhoetAl94PRB}
Jauho A~P, Wingreen N~S and Meir Y 1994 {\em Phys. Rev. B\/} {\bf 50} 5528

\bibitem{GalperinetAl07PRB}
Galperin M, Nitzan A and Ratner M~A 2007 {\em Phys. Rev. B\/} {\bf 76} 035301

\bibitem{MuhlbacheretAl08PRL}
Muhlbacher L and Rabani E 2008 {\em Phys. Rev. Lett.\/} {\bf 100} 176403

\bibitem{ChenetAl05PRB}
Chen Z~Z, Lu R and Zhu B~F 2005 {\em Phys. Rev. B\/} {\bf 71} 165324

\bibitem{GalperinetAl06PRB}
Galperin M, Nitzan A and Ratner M~A 2006 {\em Phys. Rev. B\/} {\bf 73} 045314

\bibitem{WangetAl07PRB}
Wang R~Q, Zhou Y~Q, Wang B and Xing D~Y 2007 {\em Phys. Rev. B\/} {\bf 75}
  045318

\bibitem{PaaskeetAl05PRL}
Paaske J and Flensberg K 2005 {\em Phys. Rev. Lett.\/} {\bf 94} 176801

\bibitem{YangetAl10EPL}
Yang K~H, Wu Y~P and Zhao Y~L 2010 {\em Europhys. Lett.\/} {\bf 89} 37008

\bibitem{CostietAl94JPCM}
Costi T~A, Hewson A~C and Zlatic V 1994 {\em J. Phys.:Condens. Matter\/} {\bf
  6} 2519

\bibitem{ChietAl12JPCM}
Chi F and Dubi Y 2012 {\em J. Phys.:Condens. Matter\/} {\bf 24} 145301

\bibitem{Teradaetal10JPCM}
Terada Y, Yoshida S, Takeuchi O and Shigekawa H 2010 {\em J. Phys.: Condens.
  Matter\/} {\bf 22} 264008

\bibitem{TeradaetAl10Nature}
Terada Y, Yoshida S, Takeuchi O and Shigekawa H 2010 {\em Nat. Photon.\/} {\bf
  4} 869

\end{thebibliography}

\providecommand{\newblock}{}

\end{document}